\documentstyle[12pt]{article}

\author{R.M.Sardarli, A.P.Abdullayev, O.A.Samedov, N.A.Eubova\\
Radiation Researh Department,\\
Azerbaijan Academy of Sciences,\\
H. Javid pr. 31a, Baku-370143, Azerbaijan
\and
B.R.Gadjiyev\\
Institute of Physics,\\
Azerbaijan Academy of Sciences,\\
H. Javid pr. 33, Baku-370143, Azerbaijan}
\title{The peculiarities of phase transitions in layered \\
ferroelectrics-semiconductors $ TlIn_{1-x}Fe_xS_2 $
}
\input tcilatex

\QQQ{Language}{
American English
}

\begin{document}

\maketitle
\begin{abstract}
The results of differential-thermal analysis of various samples as well as
dielectric measurements of solid solutions $TlIn_{1-x}Fe_x^{56,57}S_2$ are
presented. It has been established that phase transitions become blurred and
shift to low temperatures with increasing the concentration of defects in
the layered structure. at some threshold value the structure loses layerity,
the crystal becomes isotropig and the distribution of defects does
homogeneous.
\end{abstract}

\section{Introduction}

According to data of neutron diffraction measurements [1] and x-ray
investigations [2] the monoclinic modification of layered ferroelectrics
-semiconductors $TlInS_2$ undergoes a sequence of structural phase
transitions with the intermediate incommensurate phase having the modulation
vector $\vec q_{inc}=(\delta ;\delta ;0.25),$ where $\delta =0.012$.

$TlInS_2$ crystals have the space group symmetry $C_{2h}^6$ [2] at the
high-symmetric phase $\left( T>220K\right) $. The ionic-covalent bonds in
layers and ones of Van-der-Waals type between layers are competive
interactions in the structure of these crystals [3]. Inproper ferroelectric
phase transitions in $TlInS_2$ are intermediate between types of mixing and
order-disorder with the characteristic value of Curie's constant $C\sim 10^3K
$ [4].

An analysis of experimental data by the temperature dependence of dielectric
constant for various samples $TlInS_2$ with monoclinic structure shows that
the $\epsilon \left( T\right) $ curves characterized for proper
ferroelectrics with incommensurate phase acquire the form characteristic for
inproper ferroelectrics with incommensurate phase (Fig.1, curves 1,2,3). The
detailed study of results of neutron diffraction measurements [1], x-ray
[2], nuclear magnetic resonance and nuclear quadrupole resonance
investigations [5,6] allows to conclude that layered compounds $TlInS_2$
depending on concentration of defects in the structure have different
character of phase transitions. Besides the considerable mobility of atoms
in an interlayer space imparts to the structure the properties of weak glass
[7].

At certain value of concentration of defects in the structure the system
losing the property of layerity becomes isotropic. The range of existence of
incommensurate phase and the position of Lifshits' points on the diagram of
states depend on concentration of defects in the structure. For the most
widespread group of samples $TlInS_2$ the dependence $\epsilon \left(
T\right) $ is characterized by the evistenge of two maxima and the value of
low temperature peak varies in the large range (Fig.1, curve 2) [8].

The anomaly shows itself on the dependence $\epsilon \left( T\right) $ in
the range of existence of incommensurate phase at $T_{ic}\simeq 204K$ in
addition to maxima at $T_i$ and $T_c$ for frequently occurred and detailly
investigated samples $TlInS_2$. Experimental investigations show that the
incommensurate phase springs up in the temperature range in which the
spontaneous polarization differs from zero [9].

In seldom occurred samples $TlInS_2$ (with large concentration of defects)
the dependence $\epsilon \left( T\right) $ (Fig.1, curve 1) [10] is typical
for isotropic inproper ferroelectrics like $Rb_2Znbr_4$ [11] . In this work
the results of differential-thermal analysis (DTA) of various samples $%
TlInS_2$ as well as dielectric measurements of $TlIn_{1-x}Fe_x^{56,57}S_2$
are presented.

\section{The Results of Differential-Thermal Analysis and Dielectric
Measurements}

Single crystals of $TlInS_2$ compound were grown by modificated Bridgman
method at small variation of growth conditions. For dielectric measurements
the samples had the form of parallelepiped with parallel edges orientated
along crystallographic directions. The silver paste was used as contacts put
on polished surfaces. The measurements of dielectric constant $\epsilon $
were carried out on the frequency $1kHz$ in the temperature range $100-300K$%
. The rate of temperature change was $0.1K/min$.

The results of DTA of various samples $TlInS_2$ are presented on Fig. 2. The
samples with one maximum on the $\epsilon (T)$ dependence show only one
endothermal effect at $1020K$ on the DTA curve stipulated by melting with
the melting heat $155.75KJ/mol$. The samples $TlInS_2$ with two maxima on
the $\epsilon (T)$ dependence have two endothermal effects at temperatures $%
920$ and $1060K$ with the melting heats $80KJ/mol$ and $125KJ/mol$,
respectively. Two endothermal effects at temperatures $920$ and $1020K$ with
the melting heats $78$ and $135KJ/mol$, respectively, are detected on the
DTA curve for samples displaying three maxima in the temperature dependence
of dielectric constant.

The results of dielectric measurements for compounds $%
TlIn_{1-x}Fe_x^{56,57}S_2$, where $x=0;0.01;0.1$, at both the regime of
cooling and the regime of heating are presented on Fig. 3. An analysis of
experimental data of temperature dependence of dielectric constant $\epsilon
(T)$ allows to conclude:

-at any $x$ the values of $\epsilon (T)$ in the regime of heating are more
than in the regime of cooling;

-at any $x$ the $\epsilon (T)$ dependence displays two maxima, and the
temperature interval between them does not almost depend on temperature;

-at increasing $x$ the temperature interval of existence of incommensurate
phase shifts to low-temperature region;

-at increasing $x$ the response of system decreases;

-at increasing $x$ the transitions at $T_i$ and $T_c$ acquire the character
of blurred phase transition;

-with increasing $x$ the $\epsilon (T)$ curve loses the shape and acquires
the form characterized for inproper ferroelectrics with incommensurate phase;

-substitution of $Fe^{57}$ on $Fe^{56}$ does not change the temperature
width of existence of incommensurate phase and shifts it to high
temperatures.

\section{Discussions}

According to data of $x$-ray investigations the atoms $Tl$ in the elementary
cell of crystals $TlInS_2$ occupy the general position [2]. Besides these
atoms have high mobility in the interplane space [5]. Nuclear quadrupole
resonance investigations indicate on the anomaly at temperature $250K$
[6,12]. Evidently, the relative displacement of atoms $Tl$ in the elementary
cell occurs at $250K$ and this deformation does not change the space group
symmetry of the crystal. In addition, the deformation of the band structure
of $TlInS_2$ and levels of adhesion take place. An increase of concentration
of carriers decreases the temperature of transition to the incommensurate
phase and increases the response of this system on the external influence.
These facts explain the temperature dependence $\epsilon \left( T\right) $
in the high-symmetric phase. Although experimental data about details of
disordering in the compounds with blurred phase transitions are absent [13],
the broadening of the phase transition is general phenomenon for solid
solutions and other disordered structures.

With increasing of concentration of defects in the layered structure the
blurring of phase transitions at $T_i$ and $T_c$ and their shift to low
temperatures happen. At some threshold value the layerity of structure loses
and the crystal becomes isotropic and distribution of defects does
homogeneous. As known, if the disordering is homogeneous, a clear phase
transitions observed even in amorphous structures [14].

An increase of a depth of minimum in the temperature dependence $\epsilon
\left( T\right) $ is the result of intensification of the constant of
interaction of polarization with the amplitude of the order parameter.
Consequently, with increasing $x$ the interaction of polarization with the
amplitude of the order parameter rises while it decreases with the phase.
Incidentally, the incommensurate phase behaves inself as modulated one in
which the soliton regime is absent. The temperature hysteresis of $\epsilon
\left( T\right) $ connects with the temperature dependence of the band gap
with accuracy to the energy of adhesive level and it connects with the inner
mobility of impurities of the structure.

In the perfect layered samples $TlInS_2$ a sequence of structural phase
transitions: highsymmetrical-incommensurate $I$-incommensurate $II$%
-commensurate phase, are sprung up.

With increasing of concentration of defects in the layered structure of $%
TlInS_2$ the line of Lifshits' points on the states diagram shifts and the
incommensurate phase $II$ in dielectric measurements is not observed.

With further rise of concentration of defects in the structure the compound $%
TlInS_2$ loses the property of layerity and the sample becomes isotropic.
The temperature dependence of dielectric constant of isotropic samples at
the transition from the high-symmetric phase to the incommensurate phase
remains continuous, the incommensurate phase has not the minimum and behaves
itself as $\epsilon \sim \left( T-T_c\right) ^{-1}$. Consequently, the
dependence of physical properties of $TlInS_2$ on concentration of defects
is the result of existence of competing interactions in the structure.

The principal parameter determining the mechanism of melting is the energy
required for the rupture of bonds in the structure [14]. That is why, only
one endothermal effect shows itself in the isotropic samples of $TlInS_2$.
Layered samples $TlInS_2$, having the temperature dependence with two or
three maxima, show themselves two endothermal effects on the DTA curve.

The opposite isotropic effect connected with substitution of $Fe^{57}$ on $%
Fe^{56}$ in the $TlIn_{1-x}Fe_xS_2$ compounds is not the result of the rise
of the tunneling frequency between two positions of equilibrium. The
existence of blurred phase transition and the fact of a decrease of
temperature of phase transition with a rise of concentration of defects
testifies that the samples with $Fe^{57}$ more regulated than ones with $%
Fe^{56}$.

\begin{center}
{\bf FIGURE CAPTIONS:}
\end{center}

Fig.1. Temperature dependence of dielectric constant $\epsilon \left(
T\right) $ of crystals $TlInS_2$ (obtained from various technologic batches)
in the regime of cooling

Fig.2. The DTA curves of crystals $TlInS_2$ (obtained from various
technologic batches)

Fig.3. Temperature dependence of dielectric constant of solid solutions $%
TlIn_{1-x}Fe_xS_2$:

curves $1,2$ corresponds to the value $x=0$

curves $3,4$ - $x=0.1$ $(Fe^{56})$

curve $5$ - $x=0.01$ $(Fe^{57})$

curves $6,7$ - $x=0.1$ $(Fe^{57})$

curves $1,3,5,7$ are measured in the regime of cooling, curves $2,4,6$ -in
the regime of heating.

\end{document}